\documentclass[journal]{IEEEtran}
\IEEEoverridecommandlockouts
\usepackage{textgreek}
\usepackage{cite}
\usepackage{makecell} 

\usepackage{amsmath, microtype, graphicx} 
\usepackage{amsmath} 
\usepackage{amsmath} 

\usepackage{graphicx,color}
\usepackage{algorithm}
\usepackage{algpseudocode}
\usepackage{textcomp} 
\usepackage{bm}

\usepackage{amsmath,amssymb,amsfonts}
\usepackage{graphicx}
\usepackage{textcomp}
\usepackage{xcolor}
\usepackage[T1]{fontenc}
\usepackage{calligra}
\def\BibTeX{{\rm B\kern-.05em{\sc i\kern-.025em b}\kern-.08em
 T\kern-.1667em\lower.7ex\hbox{E}\kern-.125emX}}

 \usepackage{fancyhdr}

 \fancypagestyle{plain}{
    \fancyhf{}
    \fancyhead[L]{\footnotesize DRAFT, VOL. XX, NO. XX, AUGUST 2026}
    \fancyfoot[L]{\thepage}
    
}
\begin{document}

\title {RIS-Assisted Radar-Communication Coexistence: Detection Analysis with Channel Uncertainties}
\author{Rawan~Derbas, 
 Shimaa~Naser,~\IEEEmembership{Member,~IEEE,}
  Hamad~Yahya,~\IEEEmembership{Member,~IEEE,}
  Sanjeev~Gurugopinath,~\IEEEmembership{Senior Member,~IEEE,}  
 Paschalis~C.~Sofotasios,~\IEEEmembership{Senior Member,~IEEE,}
 Sami~Muhaidat,~\IEEEmembership{Senior Member,~IEEE}
 
 \thanks{R. Derbas, and H. Yahya are with the Department of Computer and Information Engineering, Khalifa University, Abu Dhabi, UAE (e-mails: \{rawan.derbas, hamad.myahya\}@ku.ac.ae).}
\thanks{S. Naser was with Center for Advanced Intelligent Systems, Khalifa University, Abu Dhabi 127788, UAE. She is now with the Faculty of Engineering Technology and Science, Higher Colleges of Technology, 25026, Abu Dhabi, UAE (email: snaser@hct.ac.ae).}
\thanks{S. Gurugopinath is with the Department of
Electronics and Communication Engineering, PES
University, Bengaluru, India (e-mail: sanjeevg@pes.edu).}
\thanks{P. C. Sofotasios is with Center for Intelligent Computing, Networks and Sensing, and Center for Advanced Intelligent Systems Khalifa University, Abu Dhabi
127788, UAE. He is also with the Department of Electrical Engineering, Tampere
University, 33101, Tampere, Finland (e-mail: p.sofotasios@ieee.org).}
\thanks{S. Muhaidat is with Center for Advanced Intelligent Systems, Department of Computer and Information Engineering, Khalifa University, Abu Dhabi 127788, UAE (e-mail: muhaidat@ieee.org).}
\thanks{The work of Rawan Derbas and Hamad Yahya was supported by the Faculty Startup Grant, Khalifa University of Science and Technology (KU-FSU).}
\thanks{The work of Sami Muhaidat was supported by KU Research Center for Advanced Intelligent Systems (AIS), Khalifa University of Science and Technology (KU-AIS).}
\thanks{Interested readers can find the source code for this article at https://github.com/RawanMad/RIS-RCC-channel-uncertainties.}}

\maketitle
\thispagestyle{fancy}
\begin{abstract}
Reconfigurable intelligent surfaces (RISs) have emerged as a promising technology for improving communication reliability in radar-communication coexistence (RCC) scenarios, \textcolor{black}{particularly for communication users (CUs) located inside radar exclusion zones, where only the radar is permitted to operate and the two systems remain uncoordinated.} In such settings, CUs may suffer from strong radar interference whose phase is \textcolor{black}{random and difficult to track while also facing difficulty in obtaining accurate CSI from the base station.} These challenges become even more critical in the presence of RIS phase errors, which limit the applicability of conventional coherent detection methods. Motivated by these practical limitations, this paper develops an RIS-assisted RCC framework for a CU operating in an uncoordinated RCC setting. Within this framework, we derive two practical maximum-likelihood (ML)-based detectors, both of which avoid tracking the radar interference phase: 1) a non-coherent detector that does not require instantaneous CSI and incorporates RIS phase uncertainty, and 2) a mismatched coherent detector that relies on imperfect CSI. For the non-coherent case, we derive an exact likelihood expression and a \textcolor{black}{closed-form detector in the low to moderate SINR regime}. For the imperfect CSI case, we derive the corresponding \textcolor{black}{closed-form detector in the low-to-moderate SINR regime,} and analyze performance through pairwise error probability (PEP), yielding a tractable approximation based on Gauss-Chebyshev quadrature. Numerical and analytical results show that the proposed non-coherent detector closely matches the optimal ML detector, validating its practical usefulness. The results also reveal that strong radar interference creates an interference-limited regime with a BER floor, while increasing the number of RIS reflecting elements substantially improves robustness by lowering the BER and delaying the onset of this floor. In addition, RIS phase errors and CSI mismatch degrade performance in distinct ways, but their impact can be significantly alleviated by increasing the number of reflecting elements in the RIS.
\end{abstract}

\begin{IEEEkeywords}
Interference, maximum-likelihood detector, radar-communication coexistence, reconfigurable intelligent surfaces.
\end{IEEEkeywords}

%
\IEEEpeerreviewmaketitle

%
%
%
%

\section{Introduction}

\IEEEPARstart{F}{uture} wireless communication systems are expected to support massive connectivity, high data rates, ultra-reliable operation, and emerging intelligent services that go far beyond the capabilities of fifth-generation (5G) networks \cite{giordani2020toward}. At the same time, next-generation wireless platforms are increasingly required not only to provide communication services, but also to enable accurate environment awareness and high-resolution sensing for applications such as autonomous transportation, industrial automation, extended reality, and smart healthcare. These requirements have motivated the development of integrated sensing and communication (ISAC), which aims to improve spectral and hardware efficiency by allowing radar and communication systems to share frequency resources, transmission waveforms, and hardware platforms \cite{luo2025isac}.

{ISAC systems can be realized under different levels of integration, ranging from tightly integrated dual-functional designs to more practical coexistence-based deployments. \textcolor{black}{One realization of ISAC is the radar-communication coexistence (RCC), where radar and communication systems use separate hardware platforms, waveforms, and signal processing procedures while operating over shared spectrum resources \cite{luo2025isac}. This setting is particularly attractive in practice because it does not require a major modification in their infrastructures} \cite{zhang2021enabling}. Its main challenge, however, is the \textcolor{black}{mutual interference} between the radar and communication systems, which can significantly degrade the performance of both.}

Several techniques have been proposed to mitigate interference in RCC systems, including null-space projection \cite{inter1}, joint beamforming optimization \cite{inter2}, and subcarrier allocation strategies \cite{inter3}. Another widely adopted mitigation mechanism is to impose an exclusion zone around the radar, within which communication base stations are not allowed to \textcolor{black}{be deployed} \cite{hessar2016spectrum}. Although this protects radar operation, it can severely degrade communication coverage for users located inside the exclusion zone.

A promising solution to this coverage problem is the deployment of reconfigurable intelligent surfaces (RISs) inside the exclusion zone \cite{kafafy2020stochastic}. RISs are composed of many nearly passive reflecting elements (REs) that can impose programmable phase shifts on incident signals, thereby shaping the wireless propagation environment in a favorable manner \cite{wu2021intelligent}. By appropriately configuring their phase shifts, RISs can improve signal quality, extend coverage behind blockages, and enhance communication reliability without requiring active transmission. These properties make RISs especially appealing for RCC scenarios, where direct infrastructure deployment may be restricted near radar installations.

{Despite their potential, RIS-assisted RCC systems face important practical challenges. First, communication users operating near radar systems may experience strong structured interference whose phase is generally unknown and difficult to track. Second, RIS performance is sensitive to hardware impairments, particularly RIS phase errors. Third, acquiring accurate channel state information (CSI) in RIS-assisted links is difficult because of the passive nature of the RIS and the high dimensionality of the cascaded channel \cite{wu2019intelligent}. These issues are especially critical in exclusion-zone scenarios, where users may operate under poor channel conditions and where reliable coherent reception is difficult to maintain. As a result, conventional coherent detectors that rely on accurate channel and phase knowledge may become ineffective or overly costly in practice.}

\subsection{Related Work}

{Motivated by the practical challenges of CSI acquisition and synchronization, recent studies have investigated non-coherent transmission and detection strategies for RIS-assisted wireless systems \cite{seddik2022degrees}, \cite{chen2022differential}, \cite{ino2023noncoherent}. In particular, the authors of \cite{seddik2022degrees} analyzed RIS-assisted non-coherent multi-input-multi-output (MIMO) systems and showed that the performance improves as the number of RIS REs increases. In \cite{chen2022differential}, a non-coherent demodulation scheme based on differential modulation was proposed and combined with codebook-based RIS beam training to reduce training overhead. Likewise, the authors of \cite{ino2023noncoherent} developed a non-coherent RIS-aided system using differential modulation and demonstrated robustness against channel variations with low training complexity. Although these works confirmed the potential of RIS in non-coherent communications, they did not consider interference arising from RCC scenarios.}

{On the other hand, numerous studies have addressed interference management in RCC systems. For example, the coexistence between a rotating radar and a cellular system was studied in \cite{kafafy2020stochastic}, where only a subset of base stations was allowed to transmit when they were outside the radar beam. In \cite{zheng2017adaptive}, the communication user adaptively estimated and cancelled radar interference in an uncoordinated coexistence setting. Furthermore, the authors of \cite{aydogdu2019radchat} proposed RadChat, a cooperative radar-communication protocol that coordinates orthogonal resource usage among automotive radars, while \cite{kafafy2021maximum} developed a non-orthogonal channel allocation framework to maximize the number of served vehicles under radar interference constraints. However, these RCC studies did not consider RIS-assisted propagation control, which can significantly enhance communication coverage and reliability in exclusion-zone environments.}

{Recognizing the potential of RIS, recent works have started incorporating RIS into RCC systems to alleviate interference and improve system performance \cite{wang2020ris}, \cite{shtaiwi2023sum}, \cite{he2022ris}. In \cite{wang2020ris}, a single RIS was employed to improve radar detection probability through joint optimization of the BS beamformer and RIS phase shifts. In \cite{shtaiwi2023sum}, the RIS phase shifts and BS precoding were jointly optimized to enhance the communication data rate while mitigating radar interference. Moreover, \cite{he2022ris} investigated a double-RIS-assisted RCC architecture and showed that deploying two RISs can provide better performance than a single-RIS design. Nevertheless, these studies primarily focused on transmit-side optimization, beamforming, and rate enhancement. By contrast, the design and analysis of robust communication detectors that explicitly account for unknown radar phase, RIS phase errors, and inaccurate CSI remain largely underexplored.} \textcolor{black}{ This gap is particularly crucial in scenarios where CUs are served only through RIS-assisted links and operate under strong, phase-unknown radar interference while facing challenges in obtaining accurate CSI. In such conditions, conventional coherent detectors become difficult to implement and maintain in practice.}

\subsection{Motivation and Contributions}

{The above discussion reveals a clear gap between existing RIS-assisted RCC designs and the practical receiver-side challenges encountered in realistic deployments. In exclusion-zone scenarios, the communication user may operate in the presence of strong radar interference, RIS phase uncertainty, and inaccurate CSI, while the radar interference phase remains unknown and difficult to estimate. These impairments affect the receiver in different ways: RIS phase errors degrade the coherent combining provided by the RIS, whereas imperfect CSI limits the reliability of channel-aware detection. Under such conditions, conventional coherent receivers become fragile, which motivates the need for robust detector designs that can operate effectively without requiring perfect channel knowledge or accurate interference-phase tracking.}

{Motivated by these challenges, this paper studies robust receiver design for RIS-assisted RCC in the presence of a uncoordinated RCC. We consider both a non-coherent setting with RIS phase uncertainty \textcolor{black}{with statistical CSI and a mismatched coherent setting with imperfect CSI, and we develop corresponding maximum likelihood (ML) detection and performance analysis frameworks.}}

The main contributions of this paper are summarized as follows:
\begin{itemize}
\item We develop an RIS-assisted RCC signal model for communication reception in the presence of a uncoordinated RCC, explicitly accounting for the unknown radar interference phase. Within this framework, we model two distinct practical impairments: RIS phase errors in the non-coherent setting and imperfect CSI in the mismatched coherent setting.

\item For the non-coherent setting, we derive the optimal ML expression and a \textcolor{black}{closed-form detector in the  high interference regime. This detector accounts for both ideal RIS phase configuration and RIS phase-error conditions, thereby explicitly quantifying the attenuation caused by phase uncertainty. In addition, we derive the conditional pairwise error probability (PEP) for the low-to-moderate SINR detector.}

\item For the imperfect CSI setting, we derive the mismatched ML detector and analyze its error performance under unknown radar phase. In particular, we derive the PEP and use it to obtain both a
conventional union bound and a nearest-neighbor union bound (NNUB) on
the bit error rate (BER), where the resulting integrals are evaluated via Gauss-Chebyshev Quadrature (GCQ).

\item We validate the developed analysis through Monte Carlo simulations and show that the proposed non-coherent detector closely matches the optimal ML detector over a broad range of operating conditions, thereby confirming its practical usefulness with lower implementation complexity.

\item We provide several design insights from the numerical results. Specifically, strong radar interference drives the system into an interference-limited regime that gives rise to a BER floor; increasing the number of RIS REs significantly lowers the BER and delays the onset of this floor; RIS phase errors and imperfect CSI degrade performance through different mechanisms; and, in both cases, increasing the RIS REs substantially improves robustness and interference tolerance compared with random-phase RIS configurations.
\end{itemize}

\subsection{Organization}

\textcolor{black}{The remainder of this article is organized as follows. Sec.~II presents the system model, including the RIS-assisted channel, radar interference model, and received signal model. In Sec.~III, we derive the detectors for the non-coherent setting with statistical CSI and RIS phase uncertainty, and provide the corresponding performance analysis in terms of PEP and union-bound BER. In Sec.~IV, we derive the detector for the mismatched coherent setting with imperfect CSI and provide the corresponding performance analysis in terms of PEP and union-bound BER. Sec.~V presents the numerical performance evaluation of the proposed system, while Sec.~VI concludes the article.}
\\\textit{Notations}: Matrices and vectors are denoted by an upper and lower
case boldface letters, respectively. The functions $F_X(\cdot)$ and $f_X(\cdot)$ represent the random variable's cumulative distribution function (CDF) and probability density function (PDF), respectively. Moreover, $|\cdot|$, $ \angle\cdot$, and $\mathop{\mathbb{E}} [\cdot]$ denote the absolute value, the phase, and the statistical expectation operator, respectively. $\Gamma(\cdot)$, and $\gamma(\cdot)$ represent the Gamma and lower incomplete Gamma functions, respectively, while $W_{k,r}(\cdot)$, $D_m(\cdot)$, $I_0(\cdot)$, and $Q(\cdot)$ denote, respectively, the Whittaker $W$-function, the parabolic cylinder function, the modified Bessel function of the first kind (order zero), and the one-dimensional Gaussian $Q$-function.
Also, $\mathcal{U}[a,b]$, and $\mathcal{CN}(\mu,\sigma^2)$, denotes the uniform distribution over $[a,b]$, and the circularly symmetric complex Gaussian distribution with mean $\mu$ and variance $\sigma^2$, respectively. Whereas the complex conjugate and real part of a complex number value operations are denoted as $(\cdot)^*$ and $\mathcal{R}\{\cdot\}$, respectively.
\section{System and Channel Model}
\begin{figure}
  \centering
  \includegraphics[scale=.32,
  clip,
  trim=40mm 0mm 00mm 00mm]{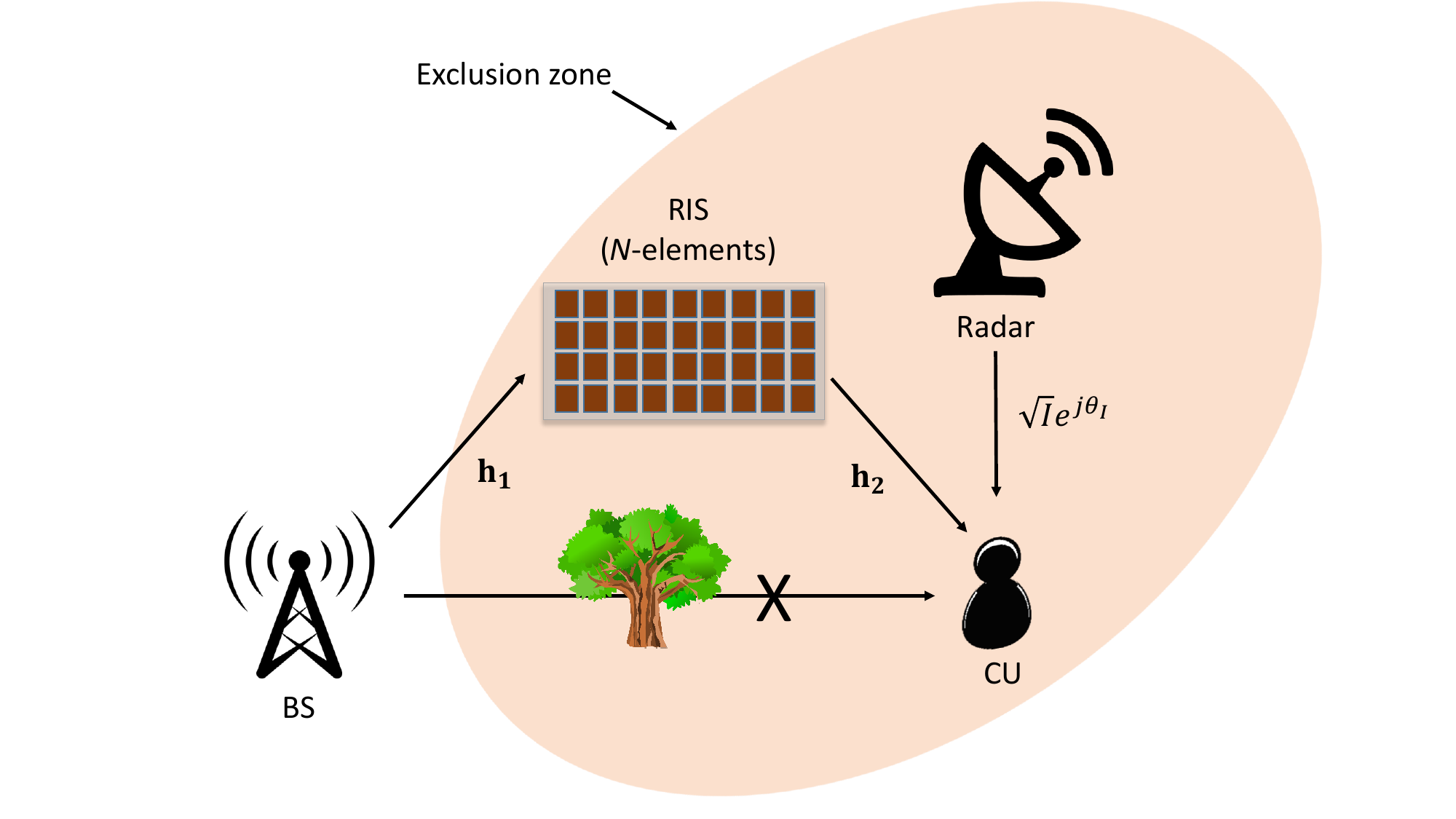} 
  \caption{Block diagram of the proposed system model.}
  \label{fig:System_MOdel}
\end{figure}
We consider an RCC network model where a directional radar is protected by an exclusion zone and a base station (BS) is allowed only outside the radar exclusion zone. \textcolor{black}{The RIS is allowed inside the exclusion zone to enhance the BS coverage of a CU located within the zone, and is assumed to be deployed outside the radar main-beam direction such that the incident radar power at the RIS, and consequently the radar-RIS-CU link, are negligible, as shown in Fig.~\ref{fig:System_MOdel}.} The RIS consists of $N$ passive REs. Each element introduces a controllable phase shift to the incoming signal and reflects it towards the CU, enhancing the received signal power. The RIS is configured by a central controller connected to the BS. The end-to-end RIS-assisted channel comprises the BS-to-RIS link and the RIS-to-CU link. We assume the direct path is blocked, and hence only the RIS-assisted link is considered. The composite channel gain is a function of the channel coefficients from the BS to each RIS element, the RIS to the CU, and the phase shifts applied by the RIS.\\

\subsection{Channel Model}
\subsubsection{Channels Through the RIS} The channel from the BS to the CU through the RIS, denoted by $B$, is
\begin{equation}
\begin{aligned}
B &= \sum_{i=1}^{N} h_{1,i}\, e^{j\phi_i}\, h_{2,i} \\
 &= \sum_{i=1}^{N} \lvert h_{1,i}\rvert \lvert h_{2,i}\rvert
 e^{j(\phi_i + \angle h_{1,i} + \angle h_{2,i})}
\end{aligned}
\end{equation}
where $h_{1,i}$ denotes the direct channel between the BS and the $i^{\text{th}}$ reflecting element of the RIS,$i=1,2,\dots,N$, and $h_{2,i}$ denotes the channel between the $i^{\text{th}}$ reflecting element of the RIS and the CU. We assume that the links $h_{1,i}$ and $h_{2,i}$ are independent and identically distributed (i.i.d.) circularly symmetric complex Gaussian (CSCG) random variables with zero mean and unit variance, i.e., $\mathcal{CN}(0, 1)$ and $h_{2,i} \sim \mathcal{CN}(0, 1)$. The vector $\bm{\phi}=[ \phi_1, \phi_2, \dots, \phi_N]^T$ is the phase shifts vector
 of the RIS.
\textcolor{black}{For the non coherent setting, we assume that the BS acquires perfect CSI via uplink training from the CU and reuses it to optimize the RIS phase shifts, exploiting channel reciprocity under time division duplex (TDD) mode \cite{larsson2014massive}. At the CU's side, the symbol detection is performed using only statistical CSI, which is relatively easy to obtain through long-term observation, varies slowly over time, and therefore requires infrequent updates, significantly reducing training overhead \cite{vila2024quadratic}.}
{\subsubsection{Radar Interference Model}
Radar systems typically transmit high-power pulses of short duration, whereas communication systems operate with much lower transmit power and usually occupy the channel continuously. As a result, from the perspective of a narrowband communication user, the radar waveform can be modeled as an additive interference term with approximately constant amplitude over the observation interval. Since radar parameters vary slowly relative to the communication symbol duration, the interference amplitude can be estimated at the receiver, for example by periodically sensing the channel prior to data transmission.
In contrast, the radar interference phase is much more difficult to track. This phase depends on the propagation delay between the radar transmitter and the communication user. Even small variations in this delay can produce large phase changes, which makes accurate phase tracking impractical. Therefore, following the commonly adopted assumption in \cite{goldsmith2005wireless} and the radar-interference model in \cite{nartasilpa2018communications,salim2017modeling}, we treat the radar phase as an unknown random variable uniformly distributed over $[0,2\pi)$.
This assumption is also supported by the results in \cite{nartasilpa2018communications,salim2017modeling}, which show that the joint amplitude-phase distribution of the radar interference can be approximated by several dominant components, each having nearly constant amplitude and uniformly distributed phase. In practice, one dominant amplitude component often captures most of the interference energy. Accordingly, in this work, the communication user models the radar interference as an additive term with effective received power $I$ and unknown phase $\theta_I \sim \mathcal{U}[0,2\pi)$. Here, $I$ already includes the effect of the radar-to-user propagation channel, including large-scale attenuation and slowly varying channel fluctuations. Thus, we do not ignore the radar-user channel; rather, we assume that its amplitude effect is captured by a dominant, approximately constant interference power over the observation interval, while the residual phase remains unknown. The known value of $I$ is explicitly incorporated into the detector design, where it enters the likelihood expressions of the optimal ML detector and the proposed detectors.}

\subsection{Received Signal Model}
The received baseband signal at the CU is given by
\begin{equation}
\label{y_first}
  y_c = \sqrt{P_t} B\, x_c + \sqrt{I} \, e^{j\theta_I} + n,
\end{equation}
where $x_c$ is the transmitted unit-energy communication symbol, $P_t$ denotes the transmitted power from the BS, and $n \sim \mathcal{CN}(0,\sigma_n^2)$ represents additive white Gaussian noise. The transmitted symbol $x_c$, and noise $n$ are assumed to be mutually uncorrelated. This model captures a practical, uncoordinated RCC scenario, where the communication system has only partial knowledge of the interference environment. This partial knowledge, particularly the uncertainty in the radar phase, motivates the adoption of non-coherent detection and robust receiver designs that can operate effectively under such conditions.

\section{Non-Coherent ML Detector}

 \subsection{Optimal ML Formulation}
 We consider ML detection at the CU for a
 non-coherent RIS-assisted system shown in Fig.~\ref{fig:System_MOdel}. The CU performs ML
 detection based on the received signal $y_c$ that is given in \eqref{y_first} as follows
 \begin{equation}
 \label{Dec_rule}
 \hat{x}_c=\mathop{\arg \max }\limits_{x_c \in \mathcal{X}} f_{Y|X}(y_c|x_c)
 \end{equation}
where $f_{Y|X}(y_c|x_c)$ is the likelihood function of the received signal, and $\mathcal{X}$ denotes the signaling constellation. {In this work, we focus on M-PSK signaling because the derivation of the proposed detector exploits its constant-modulus property, i.e., $|x_c|^2=1$ for all $x_c \in \mathcal{X}$. This enables several decision metrics to be simplified by removing symbol-energy-dependent terms. Extension to non-constant-modulus constellations is possible, but would lead to different detector expressions.}
Given the transmitted symbol $x_c$, the channel coefficient $B$, and the interference phase $\theta_I$, the conditional probability density function (PDF) of the received signal $y_c$ is expressed as
\begin{multline}
\label{step12}
f_{Y|X,B,\theta_I}(y_c|x_c, B, \theta_I) \\=\frac{1}{\pi \sigma_n^2}
\exp\left( -\frac{|y_c - \sqrt{P_t}B x_c - \sqrt{I} e^{j\theta_I}|^2}{\sigma_n^2} \right),
\end{multline}
To derive the noncoherent ML detector, it is necessary to evaluate the expectation over the PDFs of the unknown radar phase $\theta_I$ and the channel gain $B$ as follows 
\begin{multline}
f_{Y|X}(y_c|x_c) =\\ \frac{1}{2\pi} \int_0^\infty \int_0^{2\pi} f_{Y|X,B,\theta_I}(y_c|x_c, B, \theta_I)f_{\mathcal{B}}(B) d\theta_IdB.
\label{step13}
\end{multline}
This integral accounts for the uncertainty in both the radar phase and fading channel and forms the basis for the noncoherent ML detection rule.
 To account for the unknown radar phase, we first evaluate the inner integral in \eqref{step13} over \( \theta_I\):
\begin{equation}
f_{Y|X}(y_c|x_c) = \mathbb{E}_B \left[\frac{1}{2\pi} \int_0^{2\pi} f_{Y|X,B,\theta_I}(y_c|x_c, B, \theta_I) d\theta_I\right].
\label{step15}
\end{equation}
By substituting \eqref{step12} into \eqref{step15}, and expanding the term ${|y - \sqrt{P_t}B x_c - \sqrt{I} e^{j\theta_I}|^2}$, the marginal likelihood can be approximated as in \eqref{int_ini} at the top of the next page.
\begin{figure*}
\begin{multline}
\label{int_ini}
f_{Y|X}(y_c \mid x_c) 
\approx \mathbb{E}_B\!\left[
\frac{1}{\pi \sigma_n^2}
\exp\left( -\frac{|y_c - \sqrt{P_t}B x_c|^2 + I}{\sigma_n^2} \right)
\right. \\
\qquad \times \left.
\int_0^{2\pi} \frac{1}{2\pi}
\exp\left( \frac{2 |y_c - \sqrt{P_t}B x_c| \sqrt{I}}{\sigma_n^2}
\cos\left(\theta_I - \angle(y_c - \sqrt{P_t}B x_c)\right) \right)
d\theta_I
\right]
\end{multline}
\hrulefill
\end{figure*}

This integral evaluates to the modified Bessel function of the first kind and order zero, \( I_0(\cdot) \), yielding the closed-form marginal likelihood as
\begin{multline}
f_{Y|X}(y_c|x_c) = \mathbb{E}_B \!\Biggl[ \frac{1}{\pi \sigma_n^2}
\exp\left( -\frac{|y_c - \sqrt{P_t}B x_c|^2 + I}{\sigma_n^2} \right)\\ \times
I_0\left( \frac{2 |y_c - \sqrt{P_t}B x_c| \sqrt{I}}{\sigma_n^2} \right)
\Biggr].
\end{multline}

The ML detector chooses the symbol \( x_c \in \mathcal{X} \) that maximizes the marginal likelihood, thus we have 
\begin{multline}
\hat{x}_c
= \arg\max_{x_c \in \mathcal{X}} \mathbb{E}_B \!\Biggl[
\frac{1}{\pi \sigma_n^2} 
\exp\left(-\frac{|y_c - \sqrt{P_t} B x_c|^2 + I}{\sigma_n^2}\right)
\\\times I_0 \left(\frac{2 |y_c - \sqrt{P_t} B x_c| \sqrt{I}}{\sigma_n^2}\right)
\Biggr].
\end{multline}

By applying Jensen's inequality for convect maximization operation, we can write
\begin{equation}
\max_{x_c} \mathbb{E}_B[f(x_c, B)] \leq \mathbb{E}_B \left[ \max_{x_c} f(x_c, B) \right],
\end{equation}
This interchange provides an upper bound on the ML detector's marginal likelihood, facilitating analytical tractability while acknowledging the convexity of the maximization operator. Equivalently, taking the logarithms,
\begin{multline}
\hat{x}_c \approx \mathbb{E}_B
\bigg[
\arg\min_{x_c \in \mathcal{X}}
\bigg\{
\frac{|y_c - \sqrt{P_t}B x_c|^2}{\sigma_n^2}\\ - \ln I_0
\left( 
\frac{2 |y_c - \sqrt{P_t}B x_c| \sqrt{I}}{\sigma_n^2} \right)
\bigg\}
\bigg].
\end{multline}

\subsection{High-SINR Approximate ML Detector }
In the high-SINR regime, the argument inside the Bessel function becomes large, allowing the approximation
$ \ln (I_0(z)) \approx z$. Expanding $|y_c - \sqrt{P_t}B x_c|^2 = |y_c|^2 + P_t |B|^2 |x_c|^2 - 2\mathcal{R} \{ y_c^* \sqrt{P_t} B x_c \}$ and discarding terms that are {independent of $x_c$}, the lower bound can be found as 
\begin{multline}
\label{NON_ML}
\hat{x}_c \approx \arg\min_{x_c \in \mathcal{X}} \\ \left\{ \mathbb{E}_B \left[
- \mathcal{R} 
\left\{ y_c^* \sqrt{P_t} B x_c \right\}
- \sqrt{I} |y_c - \sqrt{P_t} B x_c|
\right]
\right\}.
\end{multline}

The RIS phase shift at the $i^{\text{th}}$ reflecting element is chosen as
$\phi_i = -(\angle h_{1,i}+\angle h_{2,i})$, which maximizes the received signal power at the CU. In this work, we assume that the RIS phase shifts are configured by the BS based on uplink channel estimation. That is, the CU transmits the pilots to the BS, which estimates the uplink channels and configures the RIS accordingly.

\textit{Remark 1. 
The channel gain under optimal RIS phase shift adjustment \( B= \sum_{i=1}^N |h_{1,i}||h_{2,i}| \) follows a Gamma distribution with the shape parameter \( \alpha \) and the scale parameter $\beta $ \cite{boulogeorgos2020performance}, i.e.,}
\begin{equation}
\label{pdf_non}
  f_{\mathcal{B}}(B)=\frac{B^\alpha}{\beta^{\alpha+1}\Gamma(\alpha+1)}\exp{\bigg(-\frac{B}{\beta}\bigg)},  \,\,\, B \geq 0
\end{equation}
\textit{where}
\begin{equation}
\label{alpha}
  \alpha= \frac{N\pi^2}{16-\pi^2}-1,
\end{equation}
  \textit{and} 
  \begin{equation}
  \label{beta}
  \beta=\frac{16-\pi^2}{4\pi}.
  \end{equation}
  
  \textit{The $n^{\text{th}}$ non-central moment of the above random variable can be obtained as follows}
\begin{equation}
\label{nth_moment}
  \mathbb{E}[B^n]=\mu_n=\int_0^{\infty}B^n f_{\mathcal{B}}(B)dB.
\end{equation}
\textit{Substituting \eqref{pdf_non} into \eqref{nth_moment}, and using \cite[Eq.3.381.11] { Ryzhik}, the $n^{\text{th}}$ moment can be as follows}
\begin{equation}\label{mean}\mu_n=\beta^{n}\frac{\Gamma(\alpha+n+1)}{\Gamma(\alpha+1)}.
  \end{equation}

\textcolor{black}{Building upon \eqref{NON_ML} and Remark 1, we incorporate the impact of RIS phase errors arising from RIS hardware limitations. The configured phase shifts at the RIS may deviate from the ideal optimal values, resulting in RIS phase errors. To account for this, we assume that the phase shift at the $i^{\text{th}}$
 reflecting element is given by $\hat{\phi}_i=\phi_i+\Tilde{\phi}_i$, where $\Tilde{\phi}_i$ is the
 error in the phase shift. The phase error $\Tilde{\phi}_i$ obeys independently and identically distributed (i.i.d.) random variables across $i$ having a zero mean with uniform distribution, i.e., $\Tilde{\phi}_i\sim\mathcal{U}(-\epsilon,\epsilon)$ \textcolor{black}{\cite{meanuniform,badiu2019communication}}. Thus, the channel gain becomes $ \Tilde{B}= \sum_{i=1}^N |h_{1,i}||h_{2,i}|e^{\Tilde{j\phi}_i}$, and the ML metrics in \eqref{NON_ML} can be expressed as }
\begin{multline}
\label{NON_ML_ph}
\hat{x}_c \approx \arg\min_{x_c \in \mathcal{X}} \\ \bigg\{\mathbb{E}_{\Tilde{B}} \left[
- \mathcal{R} \left\{ y_c^* \sqrt{P_t} \Tilde{B} x_c \right\}
- \sqrt{I} \big|y_c - \sqrt{P_t} \Tilde{B} x_c\big|
\right]
\bigg\}.
\end{multline}

\textit{Proposition 1. The ML rule when considering the RIS phase error at the RIS is given in \eqref{NON_ML_ph3} at the top of the next page,
where \( \eta \) is the attenuation factor, which accounts for the phase uncertainty and is given by}
\begin{equation}
  \eta = \frac{\sin(\epsilon)}{\epsilon}.
\end{equation}
\textit{Proof:} See Appendix I.
 \begin{figure*}
\begin{flalign} \begin{aligned} 
\label{NON_ML_ph3}
\hat{x}_c \approx \arg \min_{x_c \in \mathcal{X}} \bigg\{\left[ - \mathcal{R}\left\{ y_c^* \sqrt{P_t} x_c \right\} \eta \beta\frac{\Gamma(\alpha+2)}{\Gamma(\alpha+1)}
- \sqrt{I} \left| y_c - \sqrt{P_t}x_c\eta\beta\frac{\Gamma(\alpha+2)}{\Gamma(\alpha+1)} \right|
\right]\bigg\}.
\end{aligned}\end{flalign}
 \hrulefill
\end{figure*}
\\
\textcolor{black}{It is worth noting that, for a perfect RIS phase configuration, the RIS is assumed to impose ideal phase shifts without errors, leading to $\eta=1$ in \eqref{NON_ML_ph3}.}

In the low-to-moderate SINR regimes, and considering a unit modulus modulation scheme, the second part of \eqref{NON_ML_ph3}, can be ignored. Thus, the ML detector in \eqref{NON_ML_ph3} can be written as 
\begin{flalign} \begin{aligned} 
\label{NON_ML_ph_low1}
\hat{x}_c \approx \arg \min_{x_c \in \mathcal{X}} \bigg\{\left[ - \mathcal{R}\left\{ y_c^* \sqrt{P_t} x_c \right\} \eta \beta\frac{\Gamma(\alpha+2)}{\Gamma(\alpha+1)}
\right]\bigg\}.
\end{aligned}\end{flalign}
which is equivalent to the conventional Euclidean distance ML detector i.e.,   
\begin{flalign} \begin{aligned} 
\label{NON_ML_ph_low}
\hat{x}_c \approx \arg \min_{x_c \in \mathcal{X}} \bigg\{|y_c- \sqrt{P_t}  \eta \beta\frac{\Gamma(\alpha+2)}{\Gamma(\alpha+1)}x_c
|^2\bigg\}.
\end{aligned}\end{flalign}

\subsection{PEP and BER Approximation}
Capitalizing on the derived ML rule in \eqref{NON_ML_ph_low}, we apply the union bound technique as in \cite{proakisdigital} to establish an upper bound on the BER as 
 \begin{equation}
 \label{ABER_union}
 \text{P}_e\leq \frac{1}{\mathcal{M}\log_2(\mathcal{M})}\sum_{j=1}^{\mathcal{M}} \sum_{k=1,k \ne j}^{\mathcal{M}} e_j^k \Pr (x_j\rightarrow x_k ) , 
 \end{equation}
 where $ e_j^k$ represents the number of erroneous bits when decoding $x_j \text{ as } x_k$. Also, $\Pr (x_j\rightarrow x_k )$ is the event that represents the probability that, given a transmitted signal $x_j$, the CU mistakenly decides in favor of $x_k$, $j \neq k$. To this effect and capitalizing on \eqref{NON_ML_ph_low}, the expression of the PEP can be obtained as
\begin{equation}
\label{PEP_non}
\Pr(x_j \rightarrow x_k  )=(
|y_c - \sqrt{P_t}\mathcal{K} x_j|^2>
|y_c - \sqrt{P_t}\mathcal{K} x_k |^2).
\end{equation}
By expanding \eqref{PEP_non}, we have
\begin{multline}
\label{pep_ml_non}
  \Pr(x_j \rightarrow x_k  )=\Pr(\underbrace{
|\sqrt{P_t}(\Tilde{B}-\mathcal{K})x_j+\sqrt{I}e^{j\theta_I}+n|^2}_{\mathcal{J}_1}\\>
\underbrace{| \sqrt{P_t}(\Tilde{B}-\mathcal{K}) x_j+\sqrt{P_t}\mathcal{K}(x_j-x_k)+\sqrt{I}e^{j\theta_I}+n|^2}_{\mathcal{J}_2}),
\end{multline}
By expanding $\mathcal{J}_1$ and $\mathcal{J}_2$, we have
\begin{multline}
\label{J1}
   \mathcal{J}_1= P_t|(\Tilde{B}-\mathcal{K})|^2|x_j|^2+|w|^2+2\sqrt{P_t}\mathcal{R}\big\{(\Tilde{B}-\mathcal{K})^*x_j^*w\big\}
\end{multline}
\begin{multline}
\label{J2}
   \mathcal{J}_2= P_t|(\Tilde{B}-\mathcal{K})|^2+P_t\mathcal{K}^2|\Delta_{jk}|^2+|w|^2\\+2\sqrt{P_t}\mathcal{R}\big\{(\Tilde{B}-\mathcal{K})^*x_j^*w+2P_t\mathcal{R}\big\{\mathcal{K}(\Tilde{B}-\mathcal{K})^*x_j^*\Delta_{jk}\\+2\sqrt{P_t}\mathcal{R}\big\{\mathcal{K}\Delta_{jk}^*w\big\}
\end{multline}
where $w=\sqrt{I}e^{j\theta_I}+n$, and $\Delta_{jk}=x_j-x_k$. Substituting \eqref{J1} and \eqref{J2} in \eqref{pep_ml_non} and rearranging the terms, the PEP in \eqref{pep_ml_non} can be written as 
\begin{multline}
\label{pep_ml_non2}
  \Pr(x_j \rightarrow x_k  )=\Pr(2\sqrt{P_t}\,\mathcal{R}\!\left\{
\mathcal{K}^{*}\Delta_{jk}^{*}w
\right\}
<
-P_t\left|\mathcal{K}\right|^{2}
\left|\Delta_{jk}\right|^{2}
\\-2P_t\,\mathcal{R}\!\left\{
(\widetilde{B}-\mathcal{K})^{*}
x_j^{*}\mathcal{K}\Delta_{jk}
\right\}).
\end{multline}
which can be farther simplified as 
\begin{multline}
\label{PEP_ml3_non}
  \Pr(x_j \rightarrow x_k  ) =\Pr\big(\mathcal{R}\!\left\{
\mathcal{K}^{*}\Delta_{jk}^{*}n
\right\}
<
-\frac{\sqrt{P_t}}{2}
\left|\mathcal{K}\right|^{2}
\left|\Delta_{jk}\right|^{2}
\\
-\sqrt{P_t}\,\mathcal{R}\!\left\{
(\widetilde{B}-\mathcal{K})^{*}
x_j^{*}\mathcal{K}\Delta_{jk}
\right\}
-\sqrt{I}\,\mathcal{R}\!\left\{
\mathcal{K}^{*}\Delta_{jk}^{*}
e^{j\theta_I}
\right\}\big).
\end{multline}

Let \({n}^\prime =\mathcal{K}^*\Delta_{jk}^* n\). This term remains circularly symmetric because multiplying a circularly symmetric complex random variable by a deterministic phase does not change its distribution. Hence, $\mathcal{R}\{n^\prime\} \sim \mathcal{N}\!\left(0,\frac{1}{2}\mathcal{K}^2|\Delta_{jk}|^2\sigma_n^2\right)$. Moreover, the term $\sqrt{I}\,\mathcal{R}\!\left\{
\mathcal{K}^{*}\Delta_{jk}^{*}
e^{j\theta_I}
\right\}$ is equivalent to $\sqrt{I}\mathcal{K}|\Delta_{jk}|\cos(\theta_I-\angle\Delta_{jk}
)$. Also, since \(\theta_I\) is uniformly distributed, we have \(\cos(\theta_I-\angle \Delta_{jk}) \sim \cos(\theta_I)\). Thus, the conditional PEP in \eqref{PEP_ml3_non} can be rewritten as
\begin{multline}
\label{PEP_ml3_non2}
  \Pr(x_j \rightarrow x_k  ) =\mathbb{E}_{\Tilde{B},\theta_I}\bigg[Q\bigg( \frac{\sqrt{\frac{P_t}{2}}\mathcal{K} d_s-\sqrt{2I}\cos{\theta_I}}{\sqrt{\sigma_n^2}}\\-\frac{\sqrt{2P_t}\mathcal{R}\{(\Tilde{B}-\mathcal{K})x_j\Delta_{jk}^*\}}{\sqrt{\sigma_n^2}d_s}
\bigg)\bigg],
\end{multline}
where $d_s=|\Delta_{jk}|$.

\section{Mismatched Coherent Detector
}
\subsection{CSI Error Model and Effective Channel}
In this section, the RIS-assisted system model is presented under the assumption of imperfect CSI. \textcolor{black}{In RIS-aided systems with passive elements, the BS can acquire individual CSI of the BS-RIS and RIS-user links using the dual-link, reciprocity-based estimation framework proposed in~\cite{zhou2024individual}, where the BS-RIS channel is first estimated via full-duplex uplink/downlink pilot transmission and Khatri-Rao/PARAFAC-based processing, and the RIS-user channel is subsequently estimated through a least-squares (LS) step using orthogonal user pilots and known RIS reflection patterns~\cite{zhou2024individual}. As each stage relies on noisy pilot observations and linear estimators, both links naturally suffer from estimation errors, which can be captured using the model \(h = \rho \hat{h} + \sqrt{1-\rho^{2}}\,\Delta h\) \cite{suraweera2010capacity,yang2021performance,ahn2009performance} and applied separately to the BS-RIS and RIS-user channels. Here, the parameter $\rho$ denotes the correlation coefficient, scaling from 0 to 1, indicates the CSI accuracy, while $\Delta h$ represents the channel estimation error. Although the two links are, in principle, estimated under different pilot designs and SNR conditions, we adopt a common correlation coefficient \(\rho\) for both links to represent the overall CSI quality and to keep the analysis tractable, noting that the framework can be straightforwardly extended to link-specific coefficients \(\rho_{1}, \rho_{2}\) at the cost of heavier notation.} Thus, the channel coefficients are expressed accordingly
 \begin{equation}
  {h}_{1,i} =\rho \hat{h}_{1,i} + \sqrt{1 - \rho^2} \Delta h_{1,i},
 \end{equation}
 and 
 \begin{equation}
  {h}_{2,i} =\rho \hat{h}_{2,i} + \sqrt{1 - \rho^2} \Delta h_{2,i},
 \end{equation}
where $h_{1,i}$ and $h_{2,i}$ represent the actual channel coefficients, while $\hat{h}_{1,i}$ and $\hat{h}_{2,i}$ denote their corresponding estimates available at the CU. 
We assume that the estimated channels $\hat{h}_{1,i}$ and $\hat{h}_{2,i}$, as well as the estimation errors $\Delta h_{1,i}$ and $\Delta h_{2,i}$, are mutually uncorrelated. Each of these variables is modeled as a circularly symmetric complex Gaussian (CSCG) random variable with zero mean and variances matching that of the corresponding true channels; that is, $\hat{h}_{1,i}, \Delta h_{1,i} \sim \mathcal{CN}(0, 1 )$ and $\hat{h}_{2,i}, \Delta h_{2,i} \sim \mathcal{CN}(0, 1 )$. Then, the received signal at the CU can be expressed as
\begin{align}
\label{y_impc}
y_c &= \sqrt{P_t} \sum_{i=1}^N\bigg[\left( \rho \hat{h}_{1,i} + \sqrt{1 - \rho^2} \Delta h_{1,i} \right) e^{j\phi_i} \notag \\
&\times \left( \rho \hat{h}_{2,i} + \sqrt{1 - \rho^2} \Delta h_{2,i} \right) \bigg]x_c + \sqrt{I} e^{j\theta_I} + n .
\end{align}
 Furthermore, \eqref{y_impc} can be written as \eqref{y_impc2} given at the top of the next page. 
\begin{figure*}[!t]
\begin{multline}
\label{y_impc2}
y_c = 
\underbrace{\sqrt{P_t} \sum_{i=1}^N \rho^2 \hat{h}_{1,i} \hat{h}_{2,i} \, e^{j\phi_i} x_c}_{\text{Desired signal}}  
+ \underbrace{\sqrt{I} \, e^{j\theta_I}}_{\text{Radar interference term}} + n\\
+ \underbrace{
\sqrt{P_t}\bigg(\sum_{i=1}^N \left( \rho \hat{h}_{1,i} \sqrt{1 - \rho^2} \, \Delta h_{2,i} \right) e^{j\phi_i} x_c 
+ \sum_{i=1}^N \left( \rho \hat{h}_{2,i} \sqrt{1 - \rho^2} \, \Delta h_{1,i} \right) e^{j\phi_i} x_c 
+ \sum_{i=1}^N \left( (1 - \rho^2) \, \Delta h_{1,i} \Delta h_{2,i} \right) e^{j\phi_i} x_c\bigg)
}_{\text{Noise introduced by imperfect CSI}}.
\end{multline}
\hrulefill
\end{figure*}
In order to maximize the received instantaneous SINR at the CU, we set the phase shift of the $i^{\text{th}}$ RE as 
 \begin{equation}
 \label{RIS_phase}
  \hat{\phi}_i = -(\angle \hat{h}_{1,i}+ \angle \hat{h}_{2,i}), \, 1 \leq i \leq N. 
 \end{equation}
 Substituting \eqref{RIS_phase} into \eqref{y_impc2}, we obtain the following
\begin{equation}
\label{y_impc3}
y_c = \sqrt{P_t} \rho^2 \hat{B} x_c+ \sqrt{I} \, e^{j\theta_I} +\mathcal{W},
\end{equation}

where $\hat{B}=\sum_{i=1}^N |\hat{h}_{1,i}||\hat{h}_{2,i}|$, and $\mathcal{W}$ is the overall noise at the CU, which is defined in \eqref{overall_noise} at the top of the next page.

\begin{figure*}[!t]
\begin{equation}
\label{overall_noise}
\mathcal{W}=\sum_{i=1}^N \sqrt{P_t}\bigg(\big( \rho \hat{h}_{1,i} \sqrt{1 - \rho^2} \, \Delta h_{2,i} \big) e^{j\phi_i} x_c 
+ \sum_{i=1}^N \big( \rho \hat{h}_{2,i} \sqrt{1 - \rho^2} \, \Delta h_{1,i} \big) e^{j\phi_i} x_c 
+ \sum_{i=1}^N \left( (1 - \rho^2) \, \Delta h_{1,i} \Delta h_{2,i} \right) e^{j\phi_i} x_c\bigg)+
n.
\end{equation}
\hrulefill
\end{figure*}
 According to the central limit theorem (CLT), when $N\gg1$, $\mathcal{W} \sim \mathcal{CN}(0, \sigma_{\mathcal{W}}^2)$, \cite{xiong2025hybrid} with $\sigma_{\mathcal{W}}^2=  \mathbb{E}[\mathcal{W}^2]-  \mathbb{E}[\mathcal{W}]^2$, thus
\begin{multline}
\label {overall_noise_var}
\sigma_{\mathcal{W}}^2=  P_t  \rho^2 (1 - \rho^2)\sum_{i=1}^N \big(|\hat{h}_{1,i}|^2+|\hat{h}_{2,i}|^2)    + N P_t(1 - \rho^2)^2  \\  +\sigma_n^2, 
\end{multline}

\textcolor{black}{The previously derived signal model, namely \eqref{y_impc3} model, accounts for imperfect estimation of the channel gain and radar interference, while the RIS phase shifts are assumed to be known without error.
These impairments limit the applicability of conventional coherent detection schemes, which rely on accurate knowledge of both the channel and interference phase. Motivated by this, the following section develops a tailored detector suited for model \eqref{y_impc3}.}
\subsection{Mismatched ML Detector}
 Given the received signal in \eqref{y_impc3}, and the ML decision rule in \eqref{Dec_rule}, the likelihood of the received signal is given by 
 \begin{multline}
f_{Y|X,\hat{B},\theta_I}(y_c|x_c, \hat{B}, \theta_I) = \frac{1}{\pi \sigma_{\mathcal{W}}^2}\\ \times 
\exp\left( -\frac{|y_c - \sqrt{P_t}\rho^2\hat{B} x_c - \sqrt{I} e^{j\theta_I}|^2}{\sigma_{\mathcal{W}}^2} \right).
\end{multline}
By marginalizing over the unknown radar phase and following the same derivation as in the previous subsection, the optimal ML decoder can be expressed as follows
\begin{multline}
\label{imp_rule}
\hat{x}_c =
\arg\min_{x_c \in \mathcal{X}} \Bigg\{
\frac{|y_c - \sqrt{P_t}\rho^2\hat{B} x_c|^2}
{\sigma_{\mathcal{W}}^2} \\
- \ln I_0\left(
\frac{2 |y_c - \sqrt{P_t}\rho^2\hat{B} x_c| \sqrt{I}}
{\sigma_{\mathcal{W}}^2}
\right)
\Bigg\}.
\end{multline}
As a side note, in low-to-moderate SINR, we have $I_0(x)\approx1$, for $|x|\ll1$. Thus, the optimal ML decoder in \eqref{imp_rule} can be approximated as 
\begin{equation}
\label{low_INR_ML}
\hat{x}_c \approx 
\arg\min_{x_c \in \mathcal{X}} \left\{
{|y_c - \sqrt{P_t}\rho^2\hat{B} x_c|^2}
\right\}.
\end{equation}
It is noted from \eqref{low_INR_ML} that the ML detector is reduced to the conventional minimum Euclidean distance decoder, treating radar interference as noise.  
\subsection{PEP and BER Approximation}
The performance of the ML detector under imperfect CSI is studied in this section. Capitalizing on the derived ML rule in \eqref{low_INR_ML}, we apply the union bound technique in \eqref{ABER_union}. To this effect, the expression of PEP, conditioned on $\hat{B}$, can be obtained as
\begin{equation}
\label{PEP}
\Pr(x_j \rightarrow x_k |\hat{B})=(
|y_c - \sqrt{P_t}\rho^2\hat{B} x_j|^2>
|y_c - \sqrt{P_t}\rho^2\hat{B}x_k |^2).
\end{equation}
By expanding \eqref{PEP}, we have
\begin{multline}
\label{pep_ml}
  \Pr(x_j \rightarrow x_k |\hat{B})=\\ \Pr(
|\sqrt{I}e^{j\theta_I}+\mathcal{W}|^2>
| \sqrt{P_t}\rho^2\hat{B} (x_j-x_k)+\sqrt{I}e^{j\theta_I}+\mathcal{W}|^2),
\end{multline}
which can be further expanded as in \eqref{PEP_expand}
at the top of the next page.

\begin{figure*}
\begin{align}
\label{PEP_expand}
\Pr(x_j \rightarrow x_k |\hat{B}) 
&= \Pr\left(-2\mathcal{R}\left\{ \sqrt{P_t} \rho^2 \hat{B}^* (x_j -x_k )^* (\sqrt{I} e^{j\theta_I} + \mathcal{W}) \right\} 
> \left| \sqrt{P_t} \rho^2 \hat{B} (x_j -x_k ) \right|^2 \right) \nonumber \\
&= \Pr\left( \mathcal{R}\left\{ \sqrt{P_t} \rho^2 \hat{B}^* (x_j -x_k )^* \mathcal{W} \right\} 
< \frac{ \left| \sqrt{P_t} \rho^2 \hat{B} (x_j -x_k ) \right|^2 }{2} 
- \underbrace{\mathcal{R}\left\{ \sqrt{P_t} \rho^2 \hat{B}^* (x_j -x_k )^* \sqrt{I} e^{j\theta_I} \right\}}_{\mathcal{J}} \right)
\end{align}
\hrulefill
\end{figure*}

Let $\mathcal{W}^\prime=\mathcal{R}\{\sqrt{P_t}\rho^2\hat{B}^* (x_j-x_k)^*\mathcal{W}\},$ where $\mathcal{W}^\prime \sim \mathcal{N}(0,\frac{1}{2}P_t\rho^4|\hat{B}|^2|x_j-x_k|^2\sigma_{\mathcal{W}}^2)$. Also, $\mathcal{J}$ in \eqref{PEP_expand} is equivalent to $\sqrt{P_t}\rho^2|\hat{B}| |x_j-x_k|\sqrt{I}\cos(\theta_I-\angle(\hat{B}(x_j - x_k))
)$, thus the conditional PEP in \eqref{PEP_expand} can be rewritten as 
\begin{multline}
\label{PEP_ml2}
  \Pr(x_j \rightarrow x_k |\hat{B}) =\Pr\big(\mathcal{W}^\prime< \frac{|\sqrt{P_t}\rho^2\hat{B} (x_j-x_k)|^2}{2}\\-\sqrt{P_t}\rho^2|\hat{B}| |x_j-x_k|\sqrt{I}\cos(\theta_I-\angle\hat{B}(x_j - x_k)
)\big).
\end{multline}

The expression $\cos(\theta_I-\angle\hat{B}(x_j - x_k))$ does not depend on $\angle(x_j - x_k)$ because, for a circularly symmetric random variable $\mathcal{W}$ 
and an angle $\psi$, we have $\mathcal{W} e^{-j\psi} \sim \mathcal{W}$, and for a uniformly distributed random variable $\theta_I$ and an angle $\psi$, we have $\cos(\theta_I + \psi) \sim \cos(\theta_I)$. Therefore, \eqref{PEP_ml2} is only a function of the distance $d_s= |x_j - x_k|$.
Accordingly, the PEP in \eqref{PEP_ml2} can be written in terms of standard $Q$ function as 
\begin{multline}
\label{PEP_ml3}
  \Pr(x_j \rightarrow x_k) \\=\mathbb{E}_{\theta_I,\hat{B}}\bigg[Q\bigg( \frac{\sqrt{\frac{P_t}{2}}\rho^2|\hat{B}| d_s-\sqrt{2I}\cos{\theta_I}}{\sqrt{\sigma_{\mathcal{W}}^2}}
\bigg)\bigg].
\end{multline}

Unfortunately, because of the correlation between $|\hat{B}|$ in the numerator and the channel-dependent term  $\sum_{i=1}^N \big(|\hat{h}_{1,i}|^2+|\hat{h}_{2,i}|^2)$ embedded in ${\sigma_{\mathcal{W}}^2}$ in the denominator, deriving the exact PDF of the argument of the Q-function in \eqref{PEP_ml3} is hard to derive, and obtaining the exact PEP expression is challenging. To address this
problem, we take the following steps to find an approximate PEP \cite{Tag_Selec}. We redefine the value of ${\sigma_{\mathcal{W}}^2}$  in terms of it statistical variance i.e., 
\begin{multline}
\label {overall_noise22}
\bar{\sigma}_{\mathcal{W}}^2\approx  2NP_t \bar{\kappa}  \rho^2 (1 - \rho^2)+ N P_t(1 - \rho^2)^2 +\sigma_n^2, 
\end{multline}
where $\bar{\kappa}$ is a correction factor introduced to improve the accuracy of the variance approximation in the highly correlated regimes. 
\begin{multline}
\label{PEP_ml33}
  \Pr(x_j \rightarrow x_k) \\=\mathbb{E}_{\hat{B},\theta_I}\bigg[Q\bigg( \frac{\sqrt{\frac{P_t}{2}}\rho^2|\hat{B}| d_s-\sqrt{2I}\cos{\theta_I}}{\sqrt{\bar{\sigma}_{\mathcal{W}}^2}}
\bigg)\bigg].
\end{multline}
Next, we utilize the following approximation of the Q-function \cite{chiani2003new} 
 \begin{equation}
 Q(x) \approx \sum_{l=1}^2 c_l \exp(-\lambda_lx^2),
 \end{equation}
 where $c_1=\frac{1}{12}, \, c_2=\frac{1}{4}, \, \lambda_1=\frac{1}{2}, \, \text{and } \lambda_2=\frac{2}{3}$. Since $\hat{B}=\sum_{i=1}^N |\hat{h}_{1,i}||\hat{h}_{2,i}|$, its distribution can be expressed as in \eqref{pdf_non}. Thus, the average PEP can be expressed as 

 \begin{multline}
   \Pr(x_j \rightarrow x_k ) \approx \frac{1}{2\pi \Gamma(\alpha+1)\beta^{\alpha+1}} \sum_{l=1}^2 c_l\\ \times \int_{-\pi}^{\pi} \int_0^\infty 
\hat{B}^{\alpha } e^{-\frac{\hat{B}}{\beta} - \lambda_l ({\zeta_1 \hat{B} -\zeta_2\cos{\theta_I}})^2} \, d\hat{B} \, d\theta_I,
 \end{multline}
where $\zeta_1=\sqrt{\frac{\gamma}{2}}\rho^2d_s$, $\zeta_2=\sqrt{2\bar{I}}$, $\gamma=\frac{P_t}{\bar{\sigma}_{\mathcal{W}}^2}$, and $\bar{I}=\frac{I}{\bar{\sigma}_{\mathcal{W}}^2}$.

The PEP can be further
expressed as \begin{multline}
\label{intl333}
\Pr(x_j \rightarrow x_k ) \approx \frac{1}{2\pi \Gamma(\alpha+1)\beta^{\alpha+1}} \sum_{l=1}^2 c_l 
 \int_{-\pi}^{\pi} \int_0^\infty 
\hat{B}^{\alpha} \\ \exp\big(-\frac{\hat{B}}{\beta} 
- \lambda_l [ 
\zeta_1^2 \hat{B}^2 
- 2 \zeta_1\zeta_2 \hat{B} \cos{\theta_I} 
+ \zeta_2^2 \cos^2{\theta_I}
] \big) \, d\hat{B} \, d\theta_I.
\end{multline}
The following proposition describes the solution to \eqref{intl333}.
\\
\textit{Proposition 2. Using Gauss-Chebyshev
quadrature \cite[Eq. (25.4.38)] { abramowitz1964handbook}, the PEP of ML detector \eqref{low_INR_ML} is given by \eqref{chepychev_sol}}

\begin{align}
\label{chepychev_sol}
\Pr(x_j \rightarrow x_k ) 
&\;\approx\; \frac{1}{\mathcal{Q}\,\beta^{\alpha+1}}
\sum_{l=1}^{2} c_l \sum_{q=1}^{\mathcal{Q}} f_l\!\left(x_q\right),
\end{align}
where $\mathcal{Q}$ is the number of Gauss-Chebyshev nodes, $ x_q = \cos\left( \frac{2q - 1}{2\mathcal{Q}} \pi \right)$ and \( f_l(x) \) is defined as in \eqref{fx_whittaker} at the top of the next page.
\begin{figure*}
\begin{multline}
\label{fx_whittaker}
f_l(x_q) = 
\exp\left( -\lambda_l \zeta_2^2 x_q^2 \right) 
(2\lambda_l \zeta_1^2)^{-\frac{\alpha + 1}{2}} 2^{\frac{1}{4} - \frac{\alpha + 1}{2}} 
\exp\left( 
\frac{\left( \frac{1}{\beta} - 2\lambda_l \zeta_1 \zeta_2 x_q \right)^2}
{8 \lambda_l \zeta_1^2} 
\right) \\
\quad \times 
W_{ \frac{1}{4} - \frac{\alpha + 1}{2}, \, -\frac{1}{4} } 
\left( 
\frac{1}{2} \left( 
\frac{ \frac{1}{\beta} - 2 \lambda_l \zeta_1 \zeta_2 x_q }
{ \sqrt{2 \lambda_l \zeta_1^2} } 
\right)^2 
\right) 
\left( 
\frac{ \frac{1}{\beta} - 2 \lambda_l \zeta_1 \zeta_2 x_q }
{ \sqrt{2 \lambda_l \zeta_1^2} } 
\right)^{-1/2}
\end{multline}
\hrulefill
\end{figure*}
 \\ \textit{Proof:} See Appendix II.

At moderate-to-high effective SINR, the received signal is more likely to cross multiple pairwise decision boundaries simultaneously. Since union bound overestimates the probability of the
union of events by neglecting the correlations between the
events, it often exhibits a large gap with the actual performance. To overcome this problem, the Nearest Neighbor Union Bound (NNUB) is adopted to derive a tighter bound on the PEP of the composite signal \cite{proakisdigital}, thus the BER in \eqref{ABER_union} can be written as 
\begin{multline}
\label{NNUB}
  \text{P}_e\approx P_e^{\mathrm{NN}}
\\\approx \frac{N_x}{\log_2(\mathcal{M})}\mathbb{E}_{\theta_I,\hat{B}}\bigg[Q\bigg( \frac{\sqrt{\frac{P_t}{2}}\rho^2|\hat{B}| d_{\text{min}}-\sqrt{2I}\cos{\theta_I}}{\sqrt{\sigma_{\mathcal{W}}^2}}
\bigg)\bigg].
\end{multline}

$N_x$ represents the average number of the nearest
neighbor constellations whose Euclidean distance to the composite signal is equal to $d_{\text{min}}$. The analytical form of $\mathbb{E}_{\theta_I,\hat{B}}\!\left[Q\!\left(\cdot\right)\right]$ in \eqref{NNUB} is identical to that of
\eqref{chepychev_sol}, obtained by simply replacing the pairwise
constellation distance $d_s$ with the minimum Euclidean distance
$d_{\min}$.

\section{Numerical Results and Discussion}
In this section, we present simulation results to evaluate the performance of the proposed ML detection rules under unknown radar interference, considering system imperfections, i.e., the RIS REs' phase error and imperfect CSI. The performance of the detectors is studied versus the $\text{SNR}=\frac{1}{\sigma_n^2}$. To clarify the simulation setup, we provide Table I, which outlines different configurations.
\begin{table}[h]
\centering
\caption{Simulation Parameters}
\label{tab:sim_params}
\begin{tabular}{|l|c|}
\hline
\textbf{Parameter}   & \textbf{Value} \\ \hline
Transmit Power (\(P_t\))     & 1 W  \\ \hline
Number of RIS REs (\(N\))      & 6, 10, 12, 15    \\ \hline
Modulation Scheme        & QPSK    \\ \hline
$d_\text{min}$ & $2\sin(\frac{\pi}{4})$ \\ \hline
$N_x$ & 2 \\ \hline
The Correction Factor (\(\bar{\kappa}\))  & 0.7     \\ \hline
Radar Interference Power (\(I\))     &  5, 7 dBW \\ \hline
CSI error factor (\(\rho\))  & 0.8, 0.9   \\ \hline
The number of Gauss-Chebyshev nodes ($\mathcal{Q}$) & 10\\ \hline
\end{tabular}
\end{table}

\subsection{The Performance of the Noncoherent Detectors}
We begin by analyzing the noncoherent ML detector derived in Proposition 1. This is accomplished by numerically evaluating the optimal ML detector based on the likelihood function in \eqref{step13}, alongside the proposed ML detectors given in \eqref{NON_ML_ph_low}, and the union bound BER derived in \eqref{PEP_ml3_non2}.
\par
\begin{figure}
    \centering    \includegraphics[scale=.45]{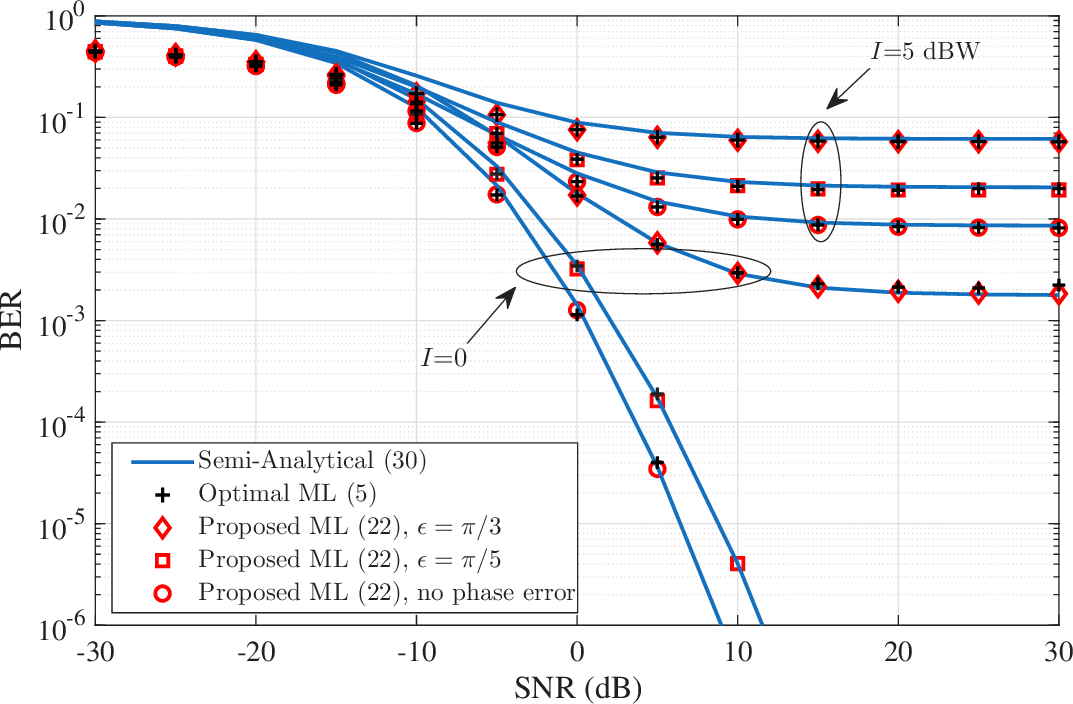}
    \caption{BER versus SNR under different radar interference power levels and RIS phase-error levels, with \(N=6\).}
    \label{fig:Fig_2}
\end{figure}

\begin{figure}
    \centering
    \includegraphics[scale=.45]{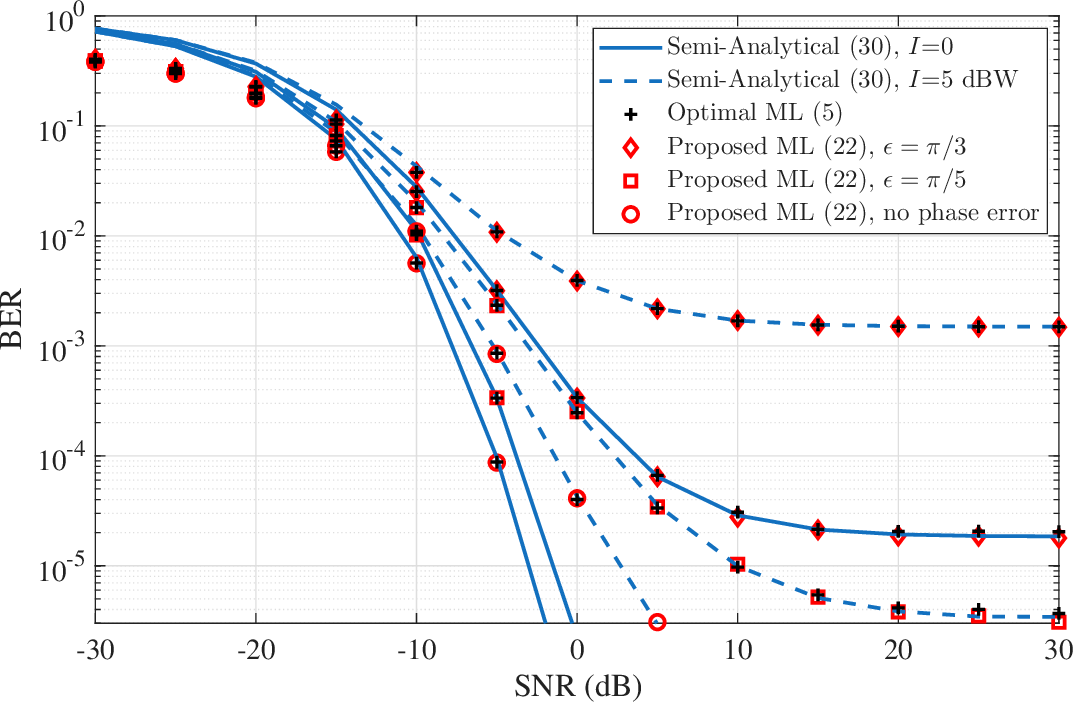}
    \caption{BER versus SNR under different radar interference power levels and RIS phase-error levels, with \(N=10\).}
    \label{fig:Fig_3}
\end{figure}

 Fig. \ref{fig:Fig_2} and Fig. \ref{fig:Fig_3} present the BER of RIS-assisted communication systems under the presence of the radar interference power, $I$, versus the SNR. Comparing the optimal detector, the proposed ML detector and the union bound for different numbers of RIS REs, we observe that the three are perfectly matched, demonstrating the accuracy and the practicality of the proposed detector under various conditions. {Fig. \ref{fig:Fig_2} illustrates the BER versus SNR for $N=6$, studied under both radar interference and no radar interference scenarios. Focusing first on the no radar interference case, the impact of RIS phase errors is evident: larger phase errors significantly degrade the performance, raising the BER floor compared to milder phase errors or the ideal no phase error scenario. For example, at $\text{SNR}=0$ dB, the BER rises from $3.4\times 10^{-3}$ to $1.7\times 10^{-2}$ when $\epsilon$ increases from $\frac{\pi}{5}$ to $\frac{\pi}{3}$ compared with $1.5\times10^{-3}$ when there is no phase error. This highlights the importance of accurate phase control in achieving the full RIS gain.
Moreover, when radar interference is introduced, the BER floor rises further due to the combined effect of interference and phase errors. For example, at $\text{SNR}=0$ dB, the BER rises from $3.9\times 10^{-2}$ to $7.6\times 10^{-2}$ when $\epsilon$ increases from $\frac{\pi}{5}$ to $\frac{\pi}{3}$ compared with $1.7\times10^{-2}$ when there is no phase error. Still, the proposed ML detector closely follows the optimal ML detector’s performance, confirming its practical utility under these challenging conditions.
\\Fig. \ref{fig:Fig_3} presents the BER versus SNR for an increased number of RIS REs, i.e., $N=10$, studied under both radar interference and no radar interference scenarios. Compared to the case $N=6$, increasing the number of RIS REs significantly enhances the BER performance. Specifically, the larger $N$ lowers the BER over the entire displayed SNR range and reduces the high-SNR floor, demonstrating a stronger mitigation of interference and phase errors. For example, at $\text{SNR}=5$ dB, and by considering the presence of radar interference and $\epsilon=\frac{\pi}{5}$, the BER reduces from $2.5\times 10^{-2}$ for $N=6$ to $3.6\times 10^{-5}$ for $N=10$ under the same conditions. These results highlight the critical role of increasing RIS REs to improve system robustness and reliability in interference-limited environments.}
\subsection{The Performance of the Mismatched ML detector}

\begin{figure}\centering
	\includegraphics[scale=.45]{ 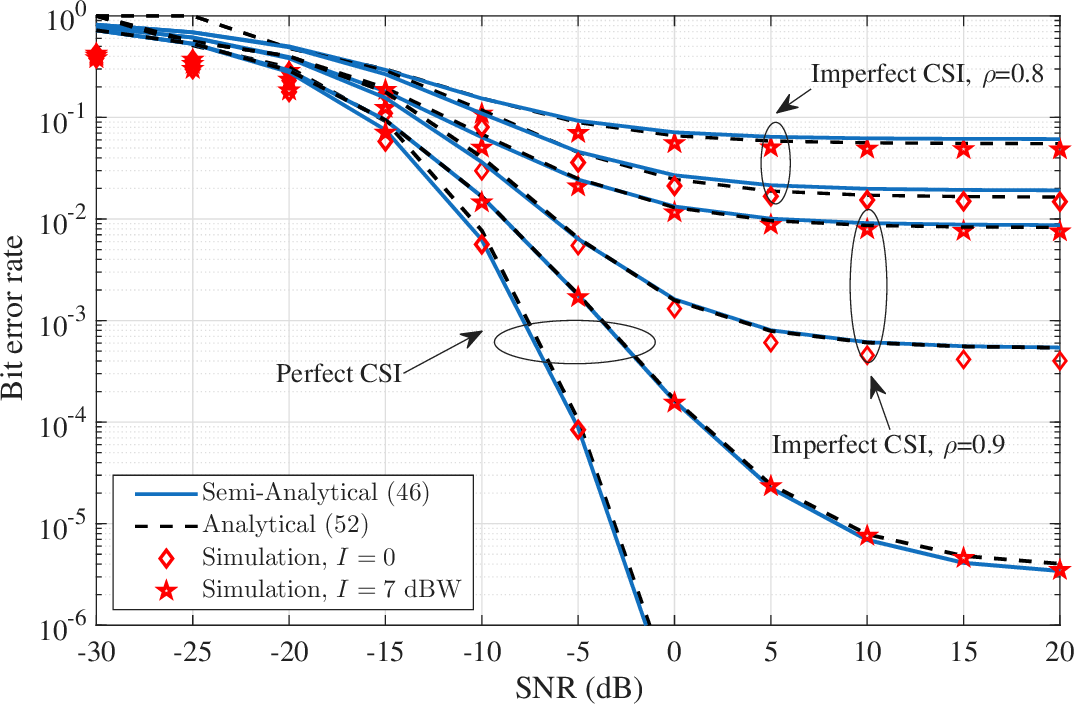}

\centering
 \caption{BER versus SNR under different radar interference power levels and CSI correlation coefficients, with \(N=12\).}
			\label{fig:Fig_4}
\end{figure}

\begin{figure}[h]\centering
	\includegraphics[scale=.45]{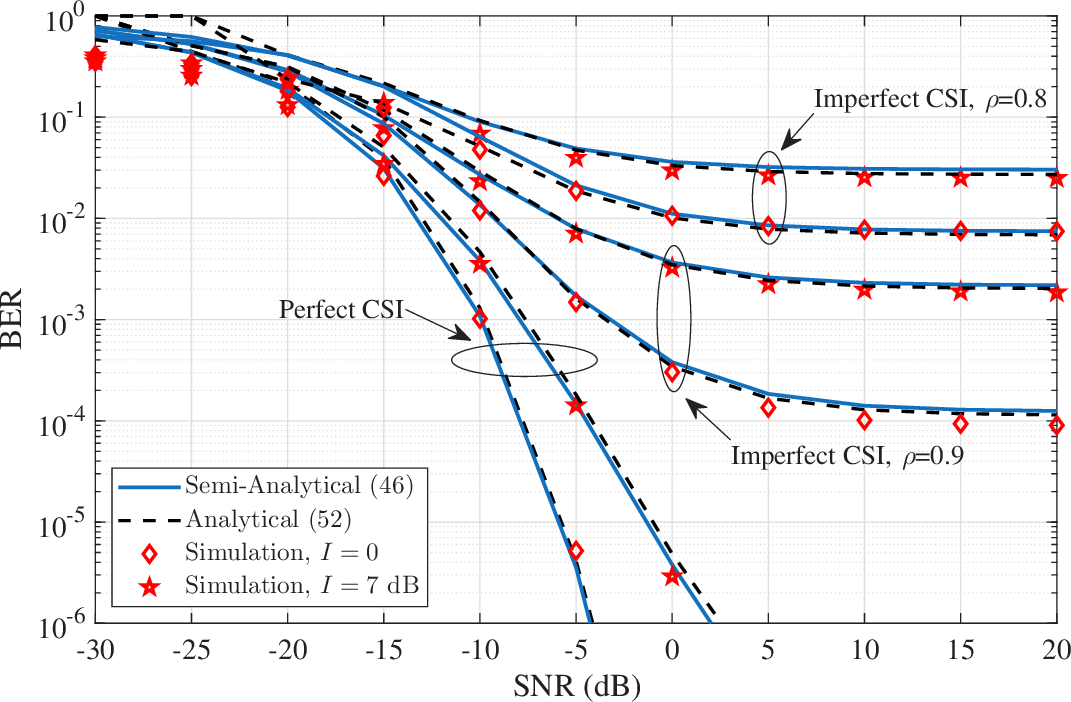}

\centering
 \caption{BER versus SNR under different radar interference power levels and CSI correlation coefficients, with \(N=15\).}
			\label{fig:Fig_5}

\end{figure}

In this subsection, we analyze the performance of the mismatched ML detector presented in \eqref{low_INR_ML}. The performance evaluation is conducted by numerically computing \eqref{ABER_union} using the semi-analytical PEP derived in \eqref{PEP_ml3} and its closed-form expression in \eqref{chepychev_sol}. The obtained simulation results are compared with the analytical union bound BER, as presented in Fig.~\ref{fig:Fig_4} and Fig.~\ref{fig:Fig_5}. It can be observed that the simulation curves closely match the analytical union bound curves obtained from \eqref{ABER_union}, confirming the accuracy of the derived expressions for both interference-free and interfered cases. Specifically, Fig.~\ref{fig:Fig_4} illustrates the impact of imperfect CSI on the BER performance of the RIS-assisted link under unknown radar phase with $N=10$. The results show that as $\rho$ decreases from $0.9$ to $0.8$, the BER floor appears at lower SNR values, around \(\text{SNR} \approx 0\,\mathrm{dB}\) for $\rho=0.8$ compared to \(\text{SNR} \approx 5\,\mathrm{dB}\) for $\rho=0.9$ under radar interference with the value $I=7$ dBW, highlighting the sensitivity of system performance to CSI quality. 

Fig.~\ref{fig:Fig_5} presents the BER versus $\text{SNR}$ with $N=15$, under various CSI correlation coefficients with different power levels of the radar interference. Compared with Fig.~\ref{fig:Fig_4}, increasing $N$ strengthens the coherent RIS-assisted desired signal, improving the effective SINR despite a fixed CSI correlation coefficient. This improvement stems from the fact that a larger number of RIS REs offers a stronger coherent combining gain, leading to a higher SINR. At high SNR, however, the system becomes interference-limited, and the BER floor is predominantly determined by $\rho$ and $I$. Lowering the CSI correlation (e.g., $\rho=0.8$) raises the error floor and makes it appear at smaller SNR, whereas with perfect CSI, the floor disappears within the plotted range, clearly revealing the RIS gain achieved with $N=15$ elements.

\par
\begin{figure}\centering
	\includegraphics[scale=.45]{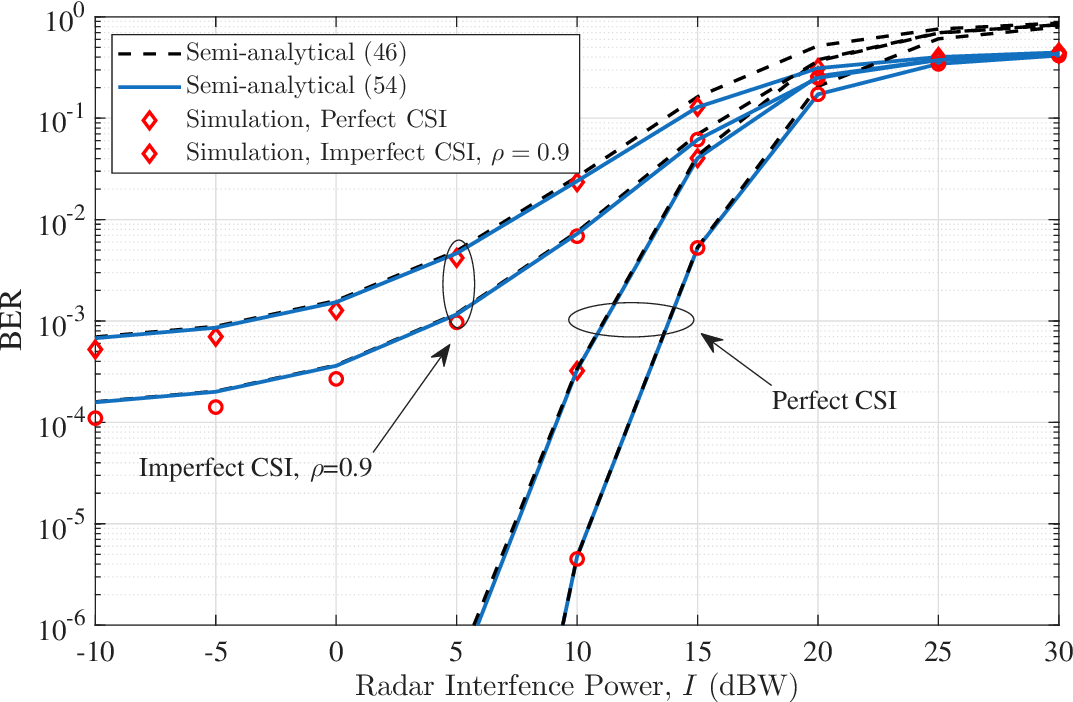}

\centering
	\caption{BER versus radar interference power $I$ at fixed $\text{SNR}=10$ dB, under different numbers of RIS REs and CSI correlation coefficients.}
			\label{fig:Fig_6}

\end{figure}

 Fig.~\ref{fig:Fig_6} shows the BER versus the radar interference power \(I\) at fixed $\text{SNR}=10$ dB, comparing different values of $N$ under perfect CSI (\(\rho=1\)) and imperfect CSI (\(\rho=0.9\)). The union bound BER using the semi-analytical PEP in \eqref{PEP_ml3} matches the simulations well. In addition, we include the NNUB, which provides a tighter approximation at high effective SINR by retaining only the dominant nearest-neighbor error events. As observed, the NNUB closely matches the simulated BER in this regime, confirming that the residual gap in the conventional union bound stems from overlapping non-dominant pairwise error events rather than from the PEP expression itself. Notably, the BER rises and eventually reaches an error floor at high interference power levels. Fortunately, adding more RIS REs enhances the system's performance. At a target $\text{BER} = 1\times 10^{-2}$, increasing $N$ from $12$ to $15$ increases the tolerable radar interference power from approximately $11$ dBW to $17$ dBW, corresponding to about a $6$ dB improvement in interference tolerance. This is mainly because adding more elements enhances the channel and boosts the signal quality.
\par
\begin{figure}\centering
	\includegraphics[scale=.45]{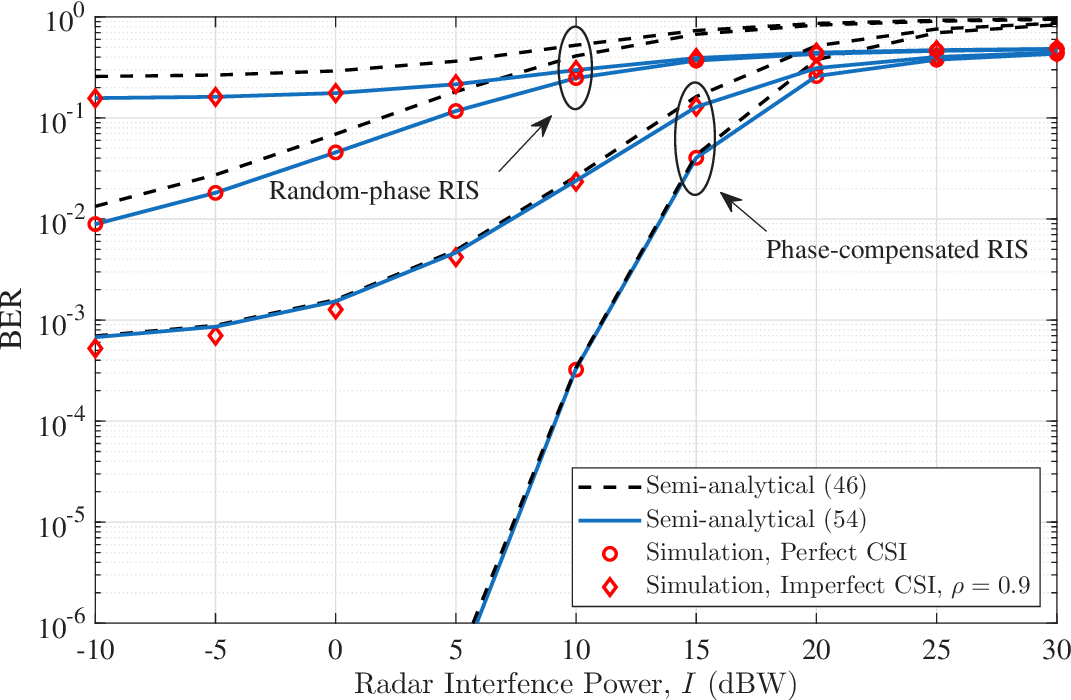}

\centering
	\caption{BER versus radar interference power $I$  under different RIS phase configurations and CSI correlation coefficients, with $\text{SNR}=10$ dB and $N=12$.}
			\label{fig:Fig_7}

\end{figure}

\textcolor{black}{Fig.~\ref{fig:Fig_7} further compares random-phase and phase-compensated RIS configurations under different CSI correlation coefficients, $\rho$. The random-phase configuration exhibits substantially worse BER performance, even under perfect CSI and at relatively low interference power levels. In contrast, phase compensation provides a considerable performance gain over a wide range of interference power levels by coherently combining the reflected signal components. Although imperfect CSI degrades the performance of the phase-compensated RIS, it still provides a pronounced improvement over the random-phase configuration. At sufficiently high interference power levels, however, the performance advantage decreases as the radar interference becomes dominant.}\\
\par
\begin{figure}\centering
	\includegraphics[scale=.45]{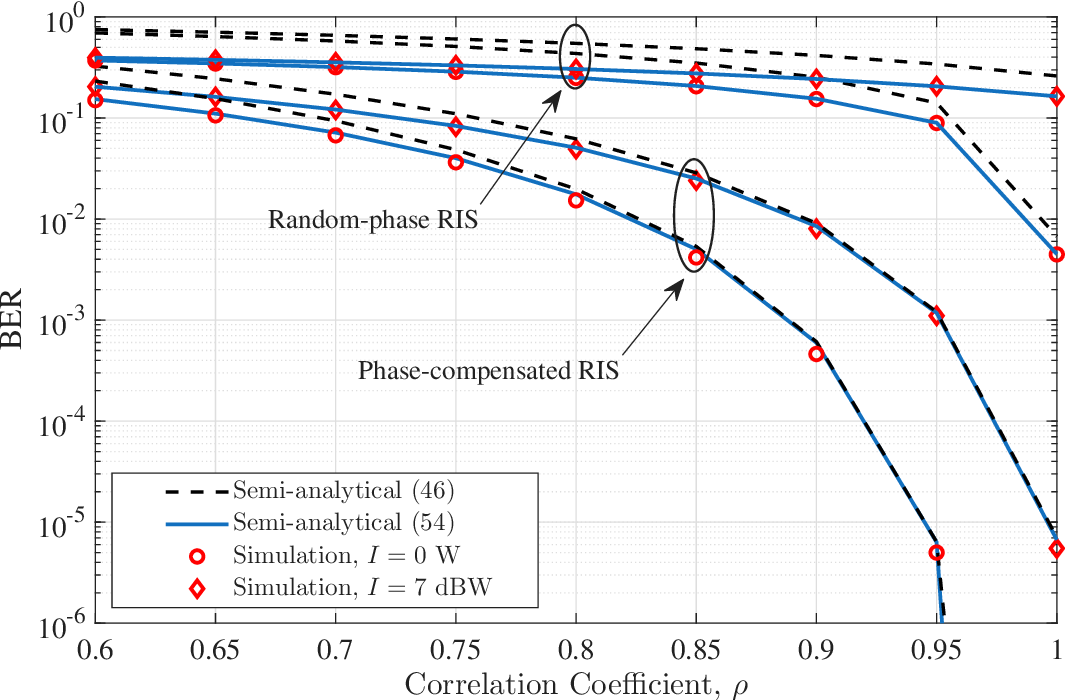}

\centering
	\caption{BER versus correlation coefficient $\rho$, under different RIS phase configurations and radar interference power levels, with $\text{SNR}=10$~dB and $N=12$.}
			\label{fig:Fig_8}

\end{figure}

 \textcolor{black}{ Fig.~\ref{fig:Fig_8} presents the BER as a function of the CSI correlation coefficient $\rho$ with and without radar interference, with $\text{SNR}=10$~dB and $N=12$. The analytical curves closely follow the simulation results, which supports the validity of the derivation. As $\rho$ increases, the BER consistently decreases, and the phase-compensated RIS maintains a clear advantage over the random-phase case. This advantage becomes more pronounced in the high-correlation regime, where the BER of the phase-compensated RIS drops sharply and reaches significantly lower values than the random-phase configuration, particularly under the higher interference power $I=7$~dBW.}


\section{Conclusion}
In this paper, we studied RIS-assisted communication in the presence of unknown-phase radar interference. In order to do so, we derived a non-coherent ML detector under unknown radar interference phase that explicitly accounts for RIS phase errors and a mismatched coherent detector for imperfect CSI. The performance is studied by analyzing PEP and union-bound BER. The proposed semi-analytical and closed-form bounds closely match simulations, indicating that the analysis accurately captures the key behaviors of the system. The results reveal several general insights. First, strong radar interference creates an interference-limited regime where the BER saturates to a floor as transmit power increases. Second, enlarging the RIS (more reflecting elements) improves reliability, lowering the BER and delaying the onset of the error floor through array/channel-hardening effects. Third, phase inaccuracies at the RIS and imperfect CSI both degrade performance by raising the floor and reducing interference tolerance. Finally, a carefully designed proposed non-coherent rule can closely track the optimal ML detector over a broad range of operating conditions, suggesting practical implementability with lower complexity.


%

\section*{APPENDIX I}
\section*{PROOF OF PROPOSITION 1}
We start with the simplified high-SINR approximation in \eqref{NON_ML_ph} of the non-coherent detection rule
\begin{multline}
\label{det_rule}
\hat{x}_c \approx
\arg\min_{x_c \in \mathcal{X}}
\Bigg[
\int_0^\infty
\Bigg(
-\mathcal{R}\left\{
y_c^* \sqrt{P_t}\,\tilde{B}x_c
\right\}
\\
-\sqrt{I}
\left|
y_c-\sqrt{P_t}\,\tilde{B}x_c
\right|
\Bigg)
f_{\mathcal{B}}(\tilde{B})\,d\tilde{B}
\Bigg].
\end{multline}

To further simplify the integral that involves  \( |y_c - \sqrt{P_t} \Tilde{B} x_c| \), we apply the triangle inequality for complex integrals \cite{rudin1987real}
\begin{multline}
\label{tri_ineq}
\int_0^\infty |y_c - \sqrt{P_t} \Tilde{B} x_c| f_{\mathcal{\Tilde{B}}}(\Tilde{B}) \, d\Tilde{B}
\geq \\ \left| \int_0^\infty (y_c - \sqrt{P_t} \Tilde{B} x_c) f_{\mathcal{\Tilde{B}}}(\Tilde{B}) \, d\Tilde{B} \right|.
\end{multline}
By using the approximation in \eqref{tri_ineq}, the decision rule in \eqref{det_rule} can be rewritten as \eqref{solll} at the top of the next page.
\begin{figure*}
\begin{flalign} \begin{aligned}
\label{solll}
\hat{x} = \arg \min_{x \in \mathcal{X}} \left[ - \mathcal{R}\left\{ y^* \sqrt{P_t} x_c \right\} \underbrace{\int_0^\infty {\Tilde{B}} f_{\mathcal{{\Tilde{B}}}}({\Tilde{B}}) d{\Tilde{B}}}_{\mathcal{I}_1}
- \sqrt{I} \left| y\underbrace{\int_0^\infty f_{\mathcal{{\Tilde{B}}}}({\Tilde{B}}) d{\Tilde{B}}}_{\mathcal{I}_2} - \sqrt{P_t}x_c\underbrace{ \int_0^\infty {\Tilde{B}} f_{\mathcal{{\Tilde{B}}}}({\Tilde{B}})}_{\mathcal{I}_1} \right|
\right].
 \end{aligned}\end{flalign}
 \hrulefill
\end{figure*}
It can be noticed that $\mathcal{I}_1$ in \eqref{solll} is the first moment (i.e., the mean) of the random variable $\Tilde{B}$. Furthermore, since \( f_{\mathcal{{\Tilde{B}}}}({\Tilde{B}}) \) is a valid PDF, the integral $\mathcal{I}_2 = \int_0^\infty f_{\mathcal{{\Tilde{B}}}}({\Tilde{B}}) \, dB = 1$
holds by definition. Thus, the ML rule in \eqref{NON_ML_ph} can be expressed as 
 \begin{multline} 
\label{NON_ML_ph2}
\hat{x}_c \\\approx \arg\min_{x_c \in \mathcal{X}} 
- \mathcal{R} \left\{ y^* \sqrt{P_t} \mathbb{E}_{\Tilde{B}}  [\Tilde{B}] x_c \right\}
- \sqrt{I} \bigg|y_c - \sqrt{P_t} \mathbb{E}_{\Tilde{B}} [\Tilde{B}]x_c\bigg|
.
\end{multline}

To evaluate the mean of the effective channel gain under RIS phase errors, recall that
$\Tilde{B}= \sum_{i=1}^N |h_{1,i}||h_{2,i}|e^{j\Tilde{\phi}_i},$
where the phase error \(\Tilde{\phi}_i\) is independent of the channel coefficients. Therefore, using the linearity of the expectation and the independence, we obtain
\begin{align}
\label{expect}
\mathbb{E}_{B,\tilde{\phi}}[\tilde{B}]
&= \mathbb{E}_{B,\tilde{\phi}}
\left[
\sum_{i=1}^N |h_{1,i}||h_{2,i}|e^{j\tilde{\phi}_i}
\right] \nonumber\\ 
&= \sum_{i=1}^N
\mathbb{E}_{B}\big[|h_{1,i}||h_{2,i}|\big]
\mathbb{E}_{\tilde{\phi}_i}\big[e^{j\tilde{\phi}_i}\big].
\end{align}

The inner expectation \( \mathbb{E}_{\Tilde{\phi}_i}[e^{j\Tilde{\phi}_i}] \) depends on the distribution of the RIS phase error \( \Tilde{\phi}_i \). When \( \Tilde{\phi}_i \sim \mathcal{U}(-\epsilon, \epsilon) \), the expectation evaluates to $\mathbb{E}_{\Tilde{\phi}_i} [e^{j\Tilde{\phi}_i}] = \frac{\sin(\epsilon)}{\epsilon}$ \cite{meanuniform}.
Thus, \eqref{expect} becomes
\begin{equation}
\mathbb{E}_{B,\Tilde{\phi}_i} \bigg[\sum_{i=1}^N |h_{1,i}||h_{2,i}| e^{j\Tilde{\phi}_i} \bigg]
= \eta \, \mathbb{E}_{B} \bigg[\sum_{i=1}^N |h_{1,i}||h_{2,i}| \bigg],
\end{equation}
where the attenuation factor \( \eta =\frac{\sin(\epsilon)}{\epsilon}\).
This result highlights the degradation in the channel gain due to RIS phase errors and incorporates the statistical characteristics of the phase noise distribution. Noting that
$
\mathbb{E}_{B}\!\left[\sum_{i=1}^N |h_{1,i}||h_{2,i}|\right] = \mu_1,$
where \(\mu_1\) is the first moment of \(B\), it follows from \eqref{mean} by setting \(n=1\) that
$
\mu_1 = \beta \frac{\Gamma(\alpha+2)}{\Gamma(\alpha+1)}.$
Substituting this result into \eqref{NON_ML_ph2}, the ML metric can be expressed as in \eqref{NON_ML_ph3}.

\section*{APPENDIX II}
\section*{PROOF OF PROPOSITION 2}
The expression in \eqref{intl333} can be further expanded and simplified as in \eqref{IntL1} at the top of the next page.
\begin{figure*}
\begin{multline}
\label{IntL1}
\Pr(x_j \rightarrow x_k ) = \frac{1}{2\pi \Gamma(\alpha+1)\beta^{\alpha+1}} \sum_{l=1}^2 c_l 
 \int_{-\pi}^{\pi} \exp\big(-\lambda_l\zeta_2^2\cos^2{\theta_I}\big)\\ \times \underbrace{\int_0^\infty 
h^{\alpha} \exp\Bigg(- \lambda_l 
\zeta_1^2 h^2
- \Big(\frac{1}{\beta} - 2 \lambda_l\zeta_1\zeta_2 \cos{\theta_I} \Big) h \Bigg) \, dh}_{\mathcal{L}_1} \, d\theta_I
\end{multline}
\hrulefill
\end{figure*}
Using \cite[Eq(3.462.1)] {Ryzhik}
to evaluate the inner integral $\mathcal{L}_1$, \eqref{IntL1} can be written as in \eqref{IntL1Solved}, where $D_{\nu}(.)$ is the parabolic cylinder function and can be expressed in terms of the Whittaker W function as given in \cite[Eq(9.240)] { Ryzhik}. Thus, \eqref{IntL1Solved} can be rewritten as \eqref{IntL1SolvedWhittaker}. Since the integral of \eqref{IntL1SolvedWhittaker} is difficult to address, we resort to the GCQ method \cite[Eq(25.4.38)] { abramowitz1964handbook} to tackle this issue.

\begin{figure*}
\begin{align}
\label{IntL1Solved}
\Pr(x_j \rightarrow x_k ) 
= \frac{1}{2\pi \beta^{\alpha+1}} \sum_{l=1}^2 c_l 
\int_{-\pi}^{\pi} \exp\left(-\lambda_l \zeta_2^2 \cos^2{\theta_I}\right) 
(2\lambda_l \zeta_1^2)^{-\frac{\alpha + 1}{2}} 
\exp\left( \frac{\left( \frac{1}{\beta} - 2\lambda_l \zeta_1\zeta_2 \cos{\theta_I} \right)^2}{8 \lambda_l \zeta_1^2} \right)\notag \\
\times D_{-(\alpha + 1)}\left( \frac{ \frac{1}{\beta} - 2 \lambda_l\zeta_1\zeta_2 \cos{\theta_I} }{ \sqrt{2 \lambda_l \zeta_1^2} } \right) 
\, d\theta_I
\end{align}
\hrulefill
\end{figure*}

\begin{figure*}[!t]
\begin{align} 
\label{IntL1SolvedWhittaker}
\Pr(x_j \rightarrow x_k ) 
&= \frac{1}{2\pi \beta^{\alpha+1}} \sum_{l=1}^2 c_l 
\int_{-\pi}^{\pi} \exp\left(-\lambda_l \zeta_2^2 \cos^2{\theta_I}\right) 
(2\lambda_l \zeta_1^2)^{-\frac{\alpha + 1}{2}} 
\exp\left( \frac{\left( \frac{1}{\beta} - 2\lambda_l \zeta_1\zeta_2 \cos{\theta_I} \right)^2}{8 \lambda_l \zeta_1^2} \right) \notag \\
&\quad \times 2^{\frac{1}{4} - \frac{\alpha + 1}{2}} 
W_{\frac{1}{4} - \frac{\alpha + 1}{2}, -\frac{1}{4}} 
\left( 
\frac{1}{2} \left( \frac{ \frac{1}{\beta} - 2 \lambda_l \zeta_1 \zeta_2 \cos{\theta_I} }{ \sqrt{2 \lambda_l \zeta_1^2} } \right)^2 
\right) 
\left( \frac{ \frac{1}{\beta} - 2 \lambda_l \zeta_1 \zeta_2 \cos{\theta_I} }{ \sqrt{2 \lambda_l \zeta_1^2} } \right)^{-1/2}
\, d\theta_I
\end{align}
\hrulefill
\end{figure*}
Let $\cos(\theta_I) = x$, we have $dx = -\sin(\theta_I)\, d\theta_I$. 
Using the trigonometric identity $\sin^2(\theta_I) = 1 - \cos^2(\theta_I)$, it follows that $\sin^2(\theta_I) = 1 - x^2$. 
Thus, we obtain
\begin{equation}
d\theta_I =
\begin{cases}
-\dfrac{dx}{\sqrt{1-x^2}}, & \theta_I \in [0,\pi], \\[2ex]
+\dfrac{dx}{\sqrt{1-x^2}}, & \theta_I \in [-\pi,0].
\end{cases}
\end{equation}
Therefore, the integral in \eqref{IntL1SolvedWhittaker} is expressed as
\begin{align}
\label{chepychev_sol1}
\Pr(x_j \rightarrow x_k ) 
&= \frac{2}{2\pi \beta^{\alpha+1}} \sum_{l=1}^2 c_l 
\int_{-1}^{1} \frac{f_l(x)}{\sqrt{1 - x^2}} dx.
\end{align}
where $f_l(x)$ is defined in \eqref{fx_whittaker}. Thus, it can be solved using GCQ as in \eqref{chepychev_sol}.

\bibliographystyle{IEEEtran}
\bibliography{bibliography.bib}
\end{document}